\documentclass[
  a4paper,
  twocolumn,
  superscriptaddress,
  nopdfoutputerror,
]{quantumarticle}

\usepackage{booktabs}
\usepackage{paralist}
\usepackage{float}
\usepackage{placeins}
\usepackage{graphicx}
\usepackage{dcolumn}
\usepackage{multirow}
\usepackage{bm}
\usepackage{microtype}
\usepackage{flushend}
\usepackage{mathtools, amsmath, braket, amsthm, amssymb, amsfonts}
\usepackage{hyperref}
\usepackage{cleveref}

\begin{document}

\title{Quantum Network-Based Prediction of Cancer Driver Genes}

\author{Patrícia Marques}
\affiliation{Instituto Superior Técnico (IST), University of Lisbon, Lisboa, Portugal}

\author{Andreas Wichert}
\affiliation{Instituto Superior Técnico (IST), University of Lisbon, Lisboa, Portugal}
\affiliation{GAIPS, INESC-ID, Porto Salvo, Portugal}

\author{Duarte Magano}
\affiliation{Faculty of Sciences, University of Porto (FCUP), Porto, Portugal}
\affiliation{21strategies GmbH, Hallbergmoos, Germany}
\author{Bruno Coutinho}
\affiliation{Instituto de Telecomunicações (IT), Lisboa, Portugal}
\affiliation{Institute of Communications and Navigation, German Aerospace Center (DLR), Weßling, Germany}

\begin{abstract}
Identification of cancer driver genes is fundamental for the development of targeted therapeutic interventions. The integration of mutational profiles with protein-protein interaction (PPI) networks offers a promising avenue for their detection~\cite{horn2018netsig, nourbakhsh2024prediction},
but scaling to large network datasets is computationally demanding. Quantum computing offers compact representations and potential complexity reductions.
Motivated by the classical method of Gumpinger \textit{et al.}~\cite{gumpinger2020prediction}, in this work we introduce a supervised quantum framework that combines mutation scores with network topology via a novel state preparation scheme, \textit{Quantum Multi-order Moment Embedding} (QMME). QMME encodes low-order statistical moments over the mutation scores of a node’s immediate and second-order neighbors, and encodes this information into quantum states. These are used as inputs to a kernel-based quantum binary classifier that discriminates known driver genes from others. 
Simulations on an empirical PPI network demonstrate competitive performance, with a 12.6\% recall gain over a classical baseline.
The pipeline performs explicit quantum state preparation and requires no classical training, enabling an efficient, nearly end-to-end quantum workflow.
A brief complexity analysis suggests the approach could achieve a quantum speedup in network-based cancer gene prediction. 
This work underscores the potential of supervised quantum graph learning frameworks to advance biological discovery.

\end{abstract}

\keywords{Binary Classification | Quantum Algorithms | Complex Networks |  Protein-Protein Interaction Networks}
\maketitle

\section{Introduction}
\noindent
A central objective in oncology is to identify and catalogue genes underlying human cancer~\cite{garraway2013lessons, vogelstein2013cancer, frampton2013development}. While genes with high mutation rates are likely drivers, most \textit{cancer driver genes} have few mutations. Methods that focus only on individual genes, for example by ranking them according to how often they are mutated, cannot reliably distinguish rare cancer driver genes from irrelevant ones~\cite{vogelstein2013cancer}.
One explanation lies in the networked nature of cellular processes: genes operate within pathways and complexes, so cancer may result from disruption at the pathway level rather than mutations in specific individual genes. Recent research has embraced this interaction-based perspective of cancer, integrating biological networks with statistical measures of gene-cancer associations to identify novel candidate cancer driver genes~\cite{horn2018netsig, reyna2018hierarchical, kovacs2019network, cheng2019network}.
In these networks, nodes correspond to genes and edges represent functional or physical interactions between. Protein-protein interaction (PPI) networks~\cite{lage2007human, li2017scored, szklarczyk2019string} are a widely used representation of gene interactions, as they integrate information from diverse data sources, tissues, and molecular processes.
\noindent Previous studies have combined PPI networks with gene-level cancer association scores, using clustering or network propagation methods.
A common choice for these gene scores is the MutSig \(P\)-value~\cite{lohr2012discovery, lawrence2014discovery}, which estimates the significance of observed mutations by considering (i) the overall mutational burden, (ii) clustering of mutations within a gene’s coding sequence, and (iii) the functional impact of mutations.
NetSig, for instance, focuses on the local neighborhood of each gene to identify cancer driver genes while correcting for node degree bias~\cite{horn2018netsig}.
A few studies have explored supervised approaches, incorporating existing knowledge of known driver genes for the prediction of new cancer genes. Gumpinger \textit{et al.} used Cancer Gene Census (CGC) annotations to train classifiers for discovering new cancer driver genes~\cite{gumpinger2020prediction}. This method models the task as a node-level supervised classification problem in a molecular interaction network, combining network topology with mutation-score features. It achieved more than a \(10\%\) improvement in area under the precision–recall curve (AUPRC) over baselines.

Such network-based approaches derive their strength from integrating weak mutation signals across interconnected genes rather than relying on strong evidence from individual genes. This shows the value of graph-structured representations as a paradigm for cancer driver gene discovery.
However, as biological networks continue to grow in size and complexity, classical methods face increasing challenges in scalability and representational efficiency. This motivates the exploration of quantum computing as an alternative framework for representing and processing such structured data, including PPI networks.
Of particular interest is quantum machine learning (QML), which has been proposed as a tool with potential for inference tasks~\cite{wittek2014quantum, schuld2015introduction, biamonte2017quantum, park2020theory}.
Recent work has begun applying quantum computing to biological network inference problems, including link prediction tasks~\cite{moutinho2023quantum, magano2023quantum, moutinho2024complexity}, demonstrating potential speedups over
classical baselines, further supporting the potential of quantum approaches in this domain. %
Inference tasks in network biology often rely on local statistical descriptors, such as neighborhood means or low-order moments, to capture structural information in graphs~\cite{gumpinger2020prediction, bi2023mm}. However, quantum embedding strategies that incorporate such statistics remain unexplored. Moreover, many existing QML approaches neglect the gate complexity of state preparation, limiting their practicality as end-to-end quantum solutions~\cite{zhao2024learning, wang2024comprehensive}.
Motivated by these gaps, we propose a quantum protocol that embeds low-order statistical moments of features in the immediate and two-hop neighborhoods into amplitude-encoded quantum states, using a novel modular unitary construction with explicit and efficient state preparation.

\begin{figure}[!b]
    \centering
    \includegraphics[width=0.95\columnwidth]{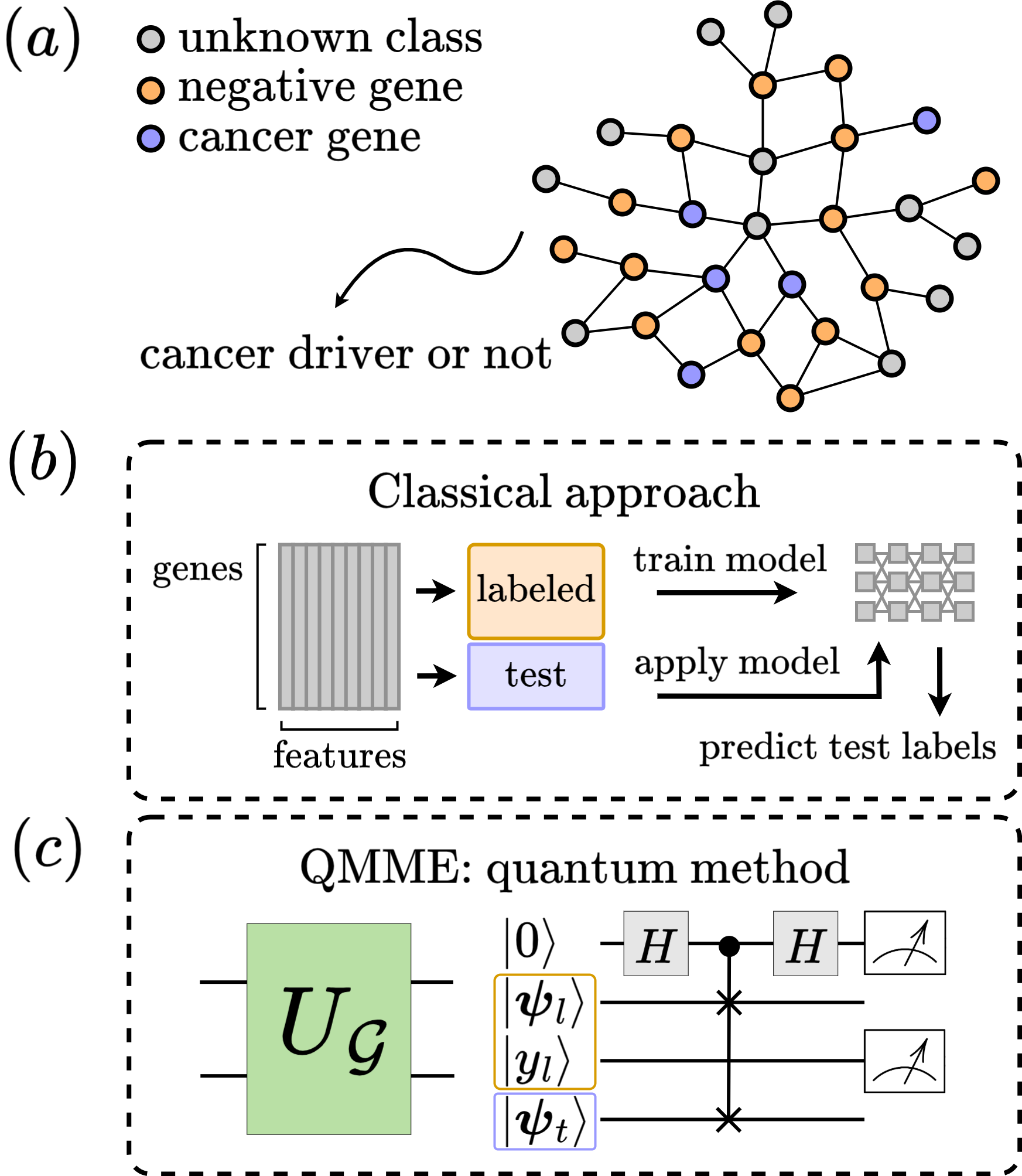} 
   \caption{\textbf{Cancer driver genes prediction.}
    \textbf{(a)} PPI network where genes are colored by class. 
    \textbf{(b)} Classical machine learning approaches: a model is trained on the labeled dataset and then used to classify the test data. 
    \textbf{(c)} Novel quantum approach: QMME (left) produces quantum embeddings, yielding the labeled state \(\ket{\psi_l}\) and test state \(\ket{\psi_t}\). The labeled state serves as a reference to classify the test state on-the-go via a swap-test (right).}  
    \label{fig:gene-net}
\end{figure}
In this work, we introduce the \textit{Quantum Multi-order Moment Embedding} (QMME) protocol, which encodes local graph statistics into quantum states for inference. We design an almost fully quantum pipeline for node-level binary classification of genes as cancer drivers.
Classification is performed using the swap-test classifier~\cite{blank2020quantum}, a non-parametric, distance-based method that employs a quantum kernel based on state similarity. We apply this pipeline to a PPI network, where nodes represent genes and features correspond to MutSig \(P\)-values, capturing mutation significance.
The entire process, from embeddings to inference, is carried out within the quantum domain, requiring no classical optimization.
It relies only on minimal data access assumptions, enabling straightforward circuit design, and involves only limited classical post-processing, inherent to the swap-test classifier.
The pipeline explicitly addresses state preparation with practical circuit construction, ensuring embeddings can be constructed with polylogarithmic qubit resources and constant-depth circuits.
We validate our approach through numerical simulations on real graphs, evaluating classification performance alongside quantum resource metrics such as circuit depth and query complexity. This work establishes a proof-of-principle for end-to-end quantum learning on graph-structured biological data and lays groundwork for future quantum models in complex network inference tasks.
\section{Problem Setup}
\noindent
Protein-protein interaction (PPI) networks model interactions between genes as nodes connected by edges, representing binding and functional associations (\Cref{fig:gene-net}a). Each gene’s association with cancer can be quantified using the MutSig \(P\)-value~\cite{lohr2012discovery, lawrence2014discovery}, which reflects the statistical significance of mutations considering mutational burden, clustering, and functional impact~\cite{gumpinger2020prediction}.
Formally, we represent a PPI network as an unweighted, undirected graph \(\mathcal{G} = (V, E, \boldsymbol{x})\) with \(|V| = N\) nodes, where \(V = [N]\) denotes the set of node indices and \(E \subseteq V \times V\) is the edge set. The connectivity of \(\mathcal{G}\) is described by its adjacency matrix \(A \in \{0,1\}^{N \times N}\), where \(A_{vu} = 1\) indicates the presence of an edge between nodes \(v\) and \(u\), and \(A_{vu} = 0\) its absence. Define the immediate neighborhood of \(v\) as
\begin{equation}
\mathcal{N}^1_v = \{ u \in V \mid A_{vu} = 1 \}.
\end{equation}
\noindent We assume \(\mathcal{N}^1_v\) is ordered by increasing node index. This allows us to define the function \(r(v, \ell)\) as
\begin{equation}
r(v, \ell) := \mathcal{N}^1_v(\ell),
\quad \ell \in \{0, \ldots, |\mathcal{N}^1_v| - 1\},
\end{equation}
\noindent which returns the index of the \(\ell\)-th neighbor of \(v\) in \(\mathcal{N}^1_v\). Equivalently, \(r(v, \ell)\) is the column index of the \(\ell\)-th nonzero entry in row \(v\) of \(A\). If \(\ell \geq |\mathcal{N}^1_v|\), i.e., the function returns a flag value \(*\) indicating no such neighbor exists.
The two-hop neighborhood \(\mathcal{N}^2_v \subseteq V\) is defined as
\begin{equation}
\mathcal{N}^2_v = \{ w \in V \mid \exists u \in V,\, A_{vu} = 1 \wedge A_{uw} = 1 \}.
\end{equation}
Each node \(v \in V\) has a scalar feature value \(x_v \in \mathbb{R}\), representing its MutSig \(P\)-value quantifying the gene’s mutation significance. The collection of all node features is denoted by the vector
\begin{equation}
\boldsymbol{x} = (x_1, x_2, \ldots, x_N) \in \mathbb{R}^N.
\label{eq:initial-features}
\end{equation}

\noindent The low-order origin (or raw) moments of a scalar random variable \( X \sim \mathcal{P} \) are defined as \(\mathbb{E}[X^i]\), where \(i\) denotes the order of the moment~\cite{shao1999mathematical}. These moments are uncentered (i.e., calculated about the origin) and capture fundamental properties of the distribution. The first four are the mean \(\mathbb{E}[X]\), the uncentered quadratic \(\mathbb{E}[X^2]\), cubic \(\mathbb{E}[X^3]\), and quartic \(\mathbb{E}[X^4]\) moments. Low-order origin moments serve as the building blocks for the more familiar summary statistics of variance, skewness, and kurtosis, which are centered or standardized variants derived from them.
Formally, we define the \emph{origin moment embedding} as the mapping
\begin{equation}
\overline{\varphi}(X) = \bigl[ \mathbb{E}[X], \;\; \mathbb{E}[X^2], \;\; \mathbb{E}[X^3], \;\; \mathbb{E}[X^4] \bigr],
\end{equation}
which collects the first four origin moments of \(X\).
Since the true distribution \(\mathcal{P}\) is generally unknown, these moments are typically estimated empirically. For a finite sample \(\boldsymbol s = [s_1, \ldots, s_k]\) of \(X\), with \(s_j \in \mathbb R\), the corresponding \emph{sample raw moment embedding} is given by
\begin{equation}
\varphi(\boldsymbol s) = \left[
\sum_{j=1}^k \frac{s_j}{k}, \quad
\sum_{j=1}^k \frac{s_j^2}{k}, \quad
\sum_{j=1}^k \frac{s_j^3}{k}, \quad
\sum_{j=1}^k \frac{s_j^4}{k}
\right].
\label{eq:sample-raw-moment-embedding}
\end{equation}
which returns the first four sample origin moments from the sample, thus, describing its distributional shape.
The hop order \(c \in \{1,2\}\) specifies the neighborhood distance from node \(v\). Given \(\boldsymbol{x}\) defined in \Cref{eq:initial-features},
\begin{equation}
\mathcal{X}^c_v = \{ x_u \in \boldsymbol{x} \mid u \in \mathcal{N}^c_v \},
\end{equation}
\noindent denotes the set of feature values of nodes at hop-distance \(c\) from \(v\), which are samples from a local distribution \(\mathcal P_v^c\).
\noindent The function \(\varphi(\cdot)\) is applied to each \(\mathcal{X}^c_v\), such that
\begin{equation}
\boldsymbol m^c_v := \varphi(\mathcal{X}^c_v) \in \mathbb{R}^4.
\end{equation}
\noindent For each vertex \(v \in V\), we define the \textit{moment-feature embedding} \(\boldsymbol m_v\in\mathbb{R}^8\) as the concatenation of \(\boldsymbol m^c_v\) from immediate and two-hop neighborhoods:
\begin{equation}
    \boldsymbol m_v = \bigl[ \boldsymbol m^1_v, \boldsymbol m^2_v \bigr] \in \mathbb{R}^8.
\label{eq:origin-moment-embedding-function}
\end{equation}

\noindent We consider a supervised binary classification task on a graph \(\mathcal{G} = (V, E, \boldsymbol{x})\), where a subset of nodes \(V_l \subseteq V\) is labeled. For these nodes, \(y_v = 1\) for \(v \in V_1 \subset V_l\) and \(y_v = 0\) for \(v \in V_0 \subset V_l\), corresponding to the positive and negative classes, respectively. This is a node-level classification task, i.e., the aim is to predict labels for individual nodes. Our goal is to predict the label \(\hat y_v\) for any node \(v \in V\), using a method that combines a \textit{quantum node embedding} unitary operator \(U_\mathcal{G}\), derived from the node features \(\boldsymbol{x}\) and the connectivity \(E\) of \(\mathcal{G}\), with a quantum binary classifier \(\mathcal{C}\). This combination is used to predict whether a node \(v\) belongs to the positive or negative class. That is,
\begin{equation}
    \mathcal{C}\left(U_\mathcal{G}\ket{v} \ket{0}^{\otimes{\mathfrak m}} \right) = \hat{y}_v,
    \label{eq:goal}
\end{equation}
\noindent where \(\hat{y}_v\in\{0,1\}\) denotes the predicted class label for node \(v\), and \(\mathfrak m\) is the number of data qubits required to encode the node embeddings.
The objective is to explicitly define \(U_\mathcal G\) as the state preparation step and design the full quantum circuit for the supervised classification of cancer driver genes. The input is a PPI network where the node features \(\boldsymbol{x}\) are per-gene MutSig \(P\)-values. For the labeled data, a set of positively labeled genes \(V_1\) is obtained from CGC data, so the positive class corresponds to cancer driver genes.

\FloatBarrier
\section{Quantum Multi-Order Moment Embedding (QMME)}
\label{sec:big-QMME}
\noindent
We base our algorithm on having query access to the graph \(\mathcal G=(V,E,\boldsymbol x)\) through the adjacency list model for bounded-degree graphs~\cite{goldreich1997property,goldreich2017introduction, ben2024symmetries}. In this input model, the graph has \(N\) vertices and bounded degree \(s=2^{\mathfrak s}\), i.e., at most \(s\) nonzero entries in each row of the adjacency matrix~\cite{lin2022lecture, camps2024explicit}. We assume there are no double edges in the network. To accommodate an attributed network, we extend this model by adding a feature query. Thus, the three available query operations are the following:
\begin{enumerate}
    \item \textbf{neighbor query:} given \(v \in V\) and \(\ell \in \mathbb Z\), returns the \(\ell\)-th neighbour of \(v\) if \(\ell < |\mathcal{N}^1_v|\), and \(*\) otherwise,
    \item \textbf{feature query:} given \(v \in V\), returns the value \(x_v\),
    \item \textbf{degree query:} given \(v \in V\), returns the degree \(k_v\).
\end{enumerate}
\noindent In the neighbor query, if the \(\ell\)-th neighbor of \(v\) does not exist (i.e., \(\ell \geq |\mathcal{N}^1_v|\)), then the flag value \(*\) is returned. A quantum equivalent of this input model can be described by defining three corresponding unitary operators \(O_L\), \(\tilde O_X\), and \(\tilde O_K\), acting on appropriate Hilbert spaces. Here, \(O_L\) provides access to a node's immediate neighborhood, \(\tilde O_X\) encodes node features, and \(\tilde O_K\) allows access to the number of \(c\)-hop neighbors. The so-called \textit{oracles} act as:
\begin{align}
    O_L &:\ket{v} \ket{\ell} \mapsto \ket{v}\ket{r(v,\ell)}, \\
    \tilde O_X &: \ket{v}\ket{0^{d'}} \mapsto \ket{v}\ket{\tilde x_v}, \\
    \tilde O_K &: \ket{v, c}\ket{0^{d'}} \mapsto \ket{v, c} \ket{\tilde k_v^c},
    \label{eq:input-model-oracles}
\end{align}
\noindent where \(v\) is the node index, \(\ell\) the neighbor index, \(\tilde x_v\) the \(d'\)-bit fixed point representation of the node feature value \(x_v \in \boldsymbol x\), and \(\tilde k_v^c\) the binary representation of the node degree \(k_v^c=|\mathcal{N}^c_v|\), i.e., the number of \(c\)-hop neighbors. 

Although oracle implementation on quantum hardware remains an active topic of research~\cite{kuklinski2024s}, our work assumes coherent access, allowing queries to be made in superposition, which is a standard assumption in the theoretical framework of quantum algorithms. We also assume that controlled versions of the oracles \(O_L\), \(\tilde O_X\), and \(\tilde O_K\) are available, together with their inverses. Since these oracles are unitary and represented by real-valued matrices, their inverses reduce to their transposes.  
In terms of query complexity, neighbor and feature queries, whether classical or quantum, are considered \textit{primitive}, and are counted as \(\mathcal{O}(1)\). By contrast, a degree query can be emulated by \(\mathcal O (\log s)\) neighbor queries using binary search~\cite{goldreich2017introduction}. To streamline second-order neighbor access, we introduced \(\tilde O_K\) as a generalization of the degree query, providing the size of a \(c\)-hop neighborhood for \(c \in \{1,2\}\). This oracle can be constructed from \(O_L\) with \(\mathcal{O}\left(\log^2 s\right)\) queries.
\noindent The goal of the embedding circuit is to encode local graph statistics into the amplitudes of quantum states. This is done using controlled single-qubit rotations, where the rotation angles are determined by values encoded in the computational basis and retrieved from the oracles. To this end, we define two rotation operators, \(F_X\) and \(F_{K^{-1}}\).
Given a node feature \(x_v \in \mathbb{R}\) with \(|x_v| \leq 1\), and its binary encoding \(\tilde{x}_v\), the \textit{feature rotation operator} \(F_X\)~\cite{lin2022lecture} is defined as
\begin{align}
    F_X &: \ket{0} \ket{\tilde{x}_v} \mapsto \left(x_v \ket{0} + \sqrt{1 - |x_v|^2} \ket{1}\right) \ket{\tilde{x}_v},
\end{align}
which corresponds to a single-qubit rotation \(R_y(\theta)\) with angle \(\theta = 2 \arcsin(x_v)\). This encodes the scalar value \(x_v\) into the amplitude of the \(\ket{0}\) component. Similarly, we define the \textit{degree-inverse rotation operator} \(F_{K^{-1}}\)~\cite{tong2021fast} as
\begin{align}
    F_{K^{-1}} &: \ket{0} \ket{\tilde{k}_v^c} \mapsto \left(\frac{1}{k_v^c} \ket{0} + \sqrt{1 - \left|\frac{1}{k_v^c}\right|^2} \ket{1}\right) \ket{\tilde{k}_v^c}.
\end{align}
These operators are implemented using multi-controlled \(R_y\) rotation gates~\cite{lin2022lecture}, where the control register holds the binary input \(\tilde{x}_v\) or \(\tilde{k}_v^c\). The rotation angles are \(\theta(\tilde{x}_v) = 2 \arcsin(x_v)\) and \(\theta(\tilde{k}_v^c) = 2 \arcsin(1 / k_v^c)\), respectively.
\noindent To amplitude-encode the moment-feature embedding \(\boldsymbol m_v\) into quantum states, we introduce two operational oracles \(O_X\) and \(O_K^{-1}\), which use the basis-encoding input oracles \(\tilde O_X\) and \(\tilde O_{K}\), along with rotation operators \(F_X\) and \(F_K^{-1}\).
The \textit{feature oracle} \(O_X\) encodes powers of the feature value \(x_v\) conditioned on the control register \(\ket{i}\). Specifically, the amplitude of the component where the ancilla register is in state \(\ket{0}^{\otimes 4}\) is mapped to \(x_v^i\), which relates to the \(i\)-th order term in the raw moment embedding function \(\varphi(\cdot)\) from \cref{eq:sample-raw-moment-embedding}.
The oracle \(O_X\) acts on a control register, where the integer \(i\) in the control register \(\ket i\) dictates how many of the ancilla qubits are rotated by \(F_X\), as follows:
\noindent 
\begin{align}
&\ket{v,i} \ket{0}^{\otimes (4+d')} 
\xrightarrow{\tilde{O}_X} 
\ket{v,i} \ket{0} \ket{\tilde{x}_v} \nonumber \\
&\xrightarrow{\mathrm{c}\text{-}F_X} 
\ket{v,i}\!\left(x_v\ket{0}\!+\!\sqrt{1\!-\!|x_v|^2}\ket{1}\right)^{\!\otimes i}\!\ket{0}^{\!\otimes (4 - i)}\ket{\tilde{x}_v} \nonumber\\
&\xrightarrow{\tilde{O}_X^{T}} 
\ket{v,i} \left( x_v \ket{0} + \sqrt{1 - |x_v|^2} \ket{1}\right)^{\otimes i} \ket{0}^{\otimes (4 - i + d')}\!.
\end{align}
\begin{figure}[!b]
    \centering
    \includegraphics[width=1\columnwidth]{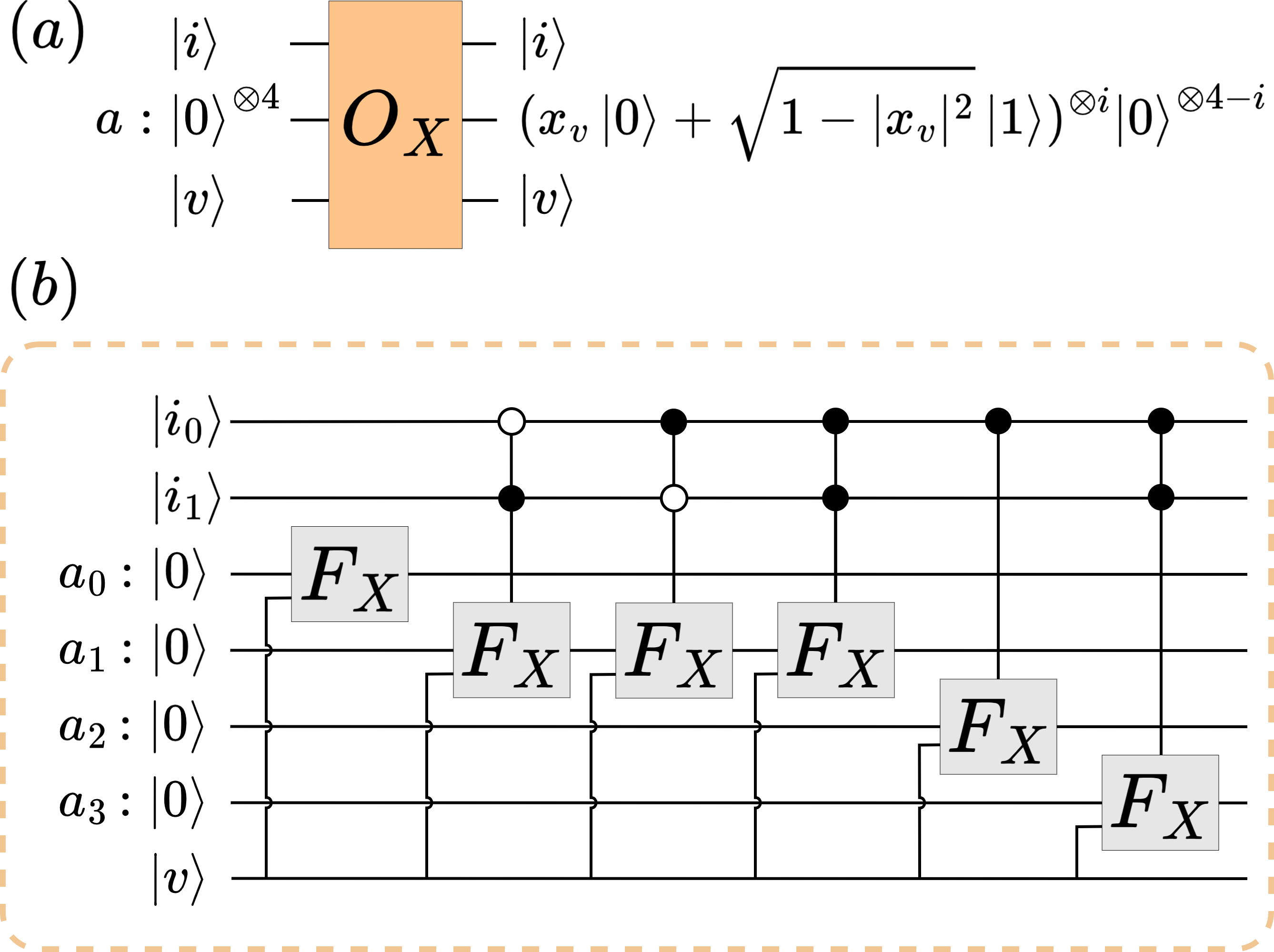}
    \caption{\textbf{Circuit of the feature oracle \(O_X\), built from controlled rotations \(F_X\).}
    \textbf{(a)} Gate representation of \(O_X\).  
    \textbf{(b)} Detailed view: a two-qubit control register \(\ket{i}\) determines how many of the four ancilla qubits undergo \(F_X\) rotations (00 → first qubit, 01 → first two, 10 → first three, 11 → all four). A \(d'\)-qubit register is used internally but omitted for clarity.}
    \label{fig:O_X}
\end{figure}
\noindent Omitting the intermediate register \(\ket{\tilde{x}_v}\), \Cref{fig:O_X} shows the circuit of oracle \(O_X\), which can be expressed as
\begin{align}
O_{X}\! :\!\ket{v,i}\!\ket{0}^{\otimes 4} \!\mapsto \!\ket{v,i}\!\left(x_v\ket{0}\!+\!\sqrt{1\!-\!|x_v|^2}\ket{1}\right)^{\!\otimes i}\!\ket{0}^{\!\otimes (4 - i)}.
\end{align}
\begin{figure*}[!t]
\includegraphics[width=\textwidth]{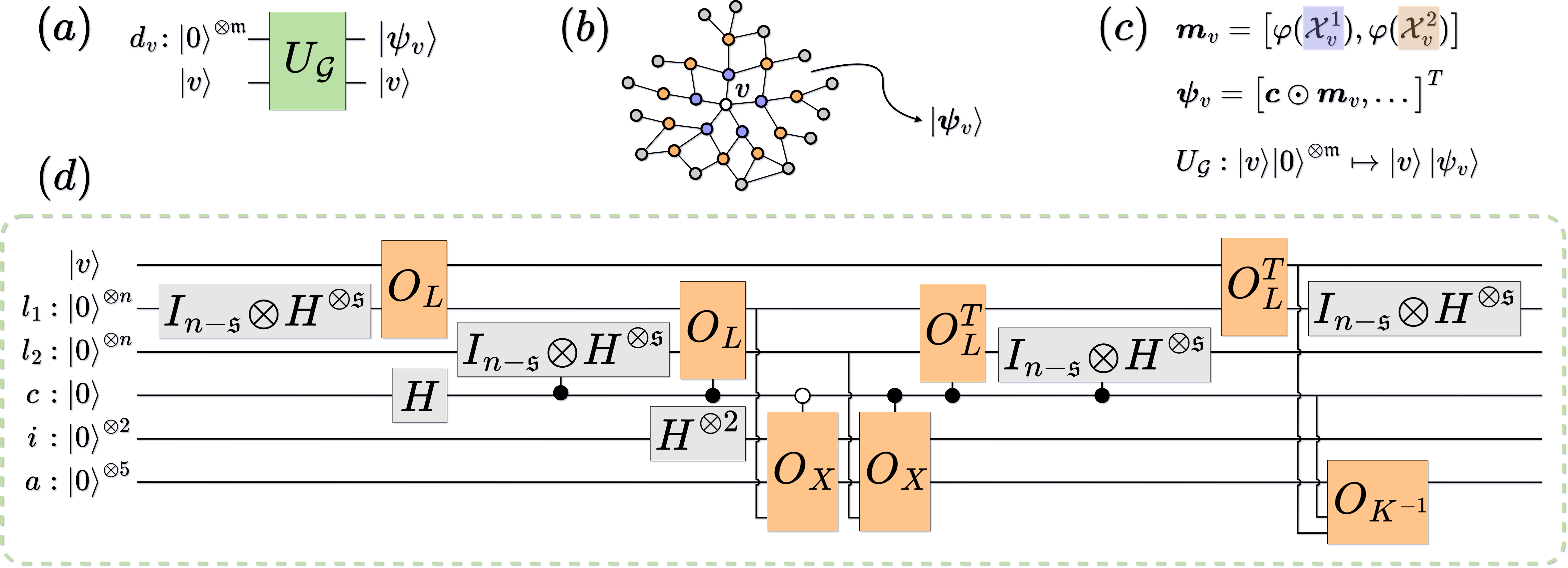}
\caption{\label{fig:qmme-schematics} \textbf{Quantum Multi-order Moment Embedding (QMME) protocol.}  
\textbf{(a)} Gate view of the unitary operator \(U_\mathcal{G}\), which maps \(\ket{v} \ket{0}^{\otimes \mathfrak{m}} \mapsto \ket{v} \ket{\boldsymbol{\psi}_v}\), amplitude-encoding low-order statistical moments of node \(v\)'s neighborhood. Here, \(\ket{v}\) encodes the node index \(v \in V\), and the \(\mathfrak{m}\)-qubit data register is initialized to zero.  
\textbf{(b)} Input PPI network \(\mathcal{G} = (V, E, \boldsymbol{x})\) with node features \(\boldsymbol{x}\) representing mutation scores. Immediate neighbors of \(v\) are shown in purple, second-hop neighbors in orange. These provide the data for the embedding \(\ket{\boldsymbol{\psi}_v} \in \mathbb{R}^{2^{\mathfrak{m}}}\).  
\textbf{(c)} The moment-feature embedding \(\boldsymbol m_v\), based on the moments \(\varphi(\cdot)\) over \(v\)'s neighborhood, is encoded as the quantum state \(\ket{\boldsymbol{\psi}_v}\).  
\textbf{(d)} Circuit-level implementation of the gate \(U_\mathcal{G}\). The \(\mathfrak{m}\)-qubit register is composed of subregisters for immediate and second-hop (\(\ell_1\),\(\ell_2\)) neighbors, a control qubit \(c\), a two-qubit register \(i\) indexing the \(i\)-th moment, and a five-qubit ancilla register \(a\). The circuit accesses the PPI network in (b) via the quantum oracles in~\cref{eq:input-model-oracles}.}
\end{figure*}
\\
\noindent
The degree-inverse oracle \(O_{K^{-1}}\) amplitude-encodes the inverse degree scaling factors \(1/k_v^c\), required to emulate the average present in the moment embedding function \(\varphi(\cdot)\), included in the moment-feature embedding \(\boldsymbol m_v\). It performs a controlled rotation that encodes the value \(1/k_v^c\) as the amplitude of the \(\ket{0}\) state of an ancilla, as follows:
\begin{equation}
\begin{aligned}
\ket{v,c} &\ket{0}^{\otimes (1+d')}
\xrightarrow{\tilde O_K} 
\ket{v,c} \ket{0} \ket{\tilde{k}_v^c} \\
&\xrightarrow{F_{K^{-1}}} 
\ket{v,c} \left( \frac{1}{k_v^c} \ket{0} + \sqrt{1 - \left| \frac{1}{k_v^c} \right|^2} \ket{1} \right) \ket{\tilde{k}_v^c} \\
&\xrightarrow{\tilde O_K^{T}} 
\ket{v,c} \left( \frac{1}{k_v^c} \ket{0} + \sqrt{1 - \left| \frac{1}{k_v^c} \right|^2} \ket{1} \right) \ket{0}^{\otimes{d'}}.
\end{aligned}
\end{equation}
\noindent As shown in \Cref{fig:qmme-schematics}, the intermediate register \(\ket{\tilde{k}_v}\) is omitted, and the oracle is represented as:
\begin{align}
O_{K^{-1}} : \ket{v,c,0} \mapsto \ket{v,c}\!\left( \frac{1}{k_v^c} \ket{0} + \sqrt{1\! - \!\left| \frac{1}{k_v^c} \right|^2} \ket{1} \right)\!.\!
\end{align}
\noindent
Together with the neighbor oracle \(O_L\) and Hadamard gates, the oracles \(O_X\) and \(O_{K^{-1}}\) form the core building blocks of the QMME circuit (depicted in \Cref{fig:qmme-schematics}).
\begin{figure*}[bht]
    \centering
    \includegraphics[width=0.7\textwidth]{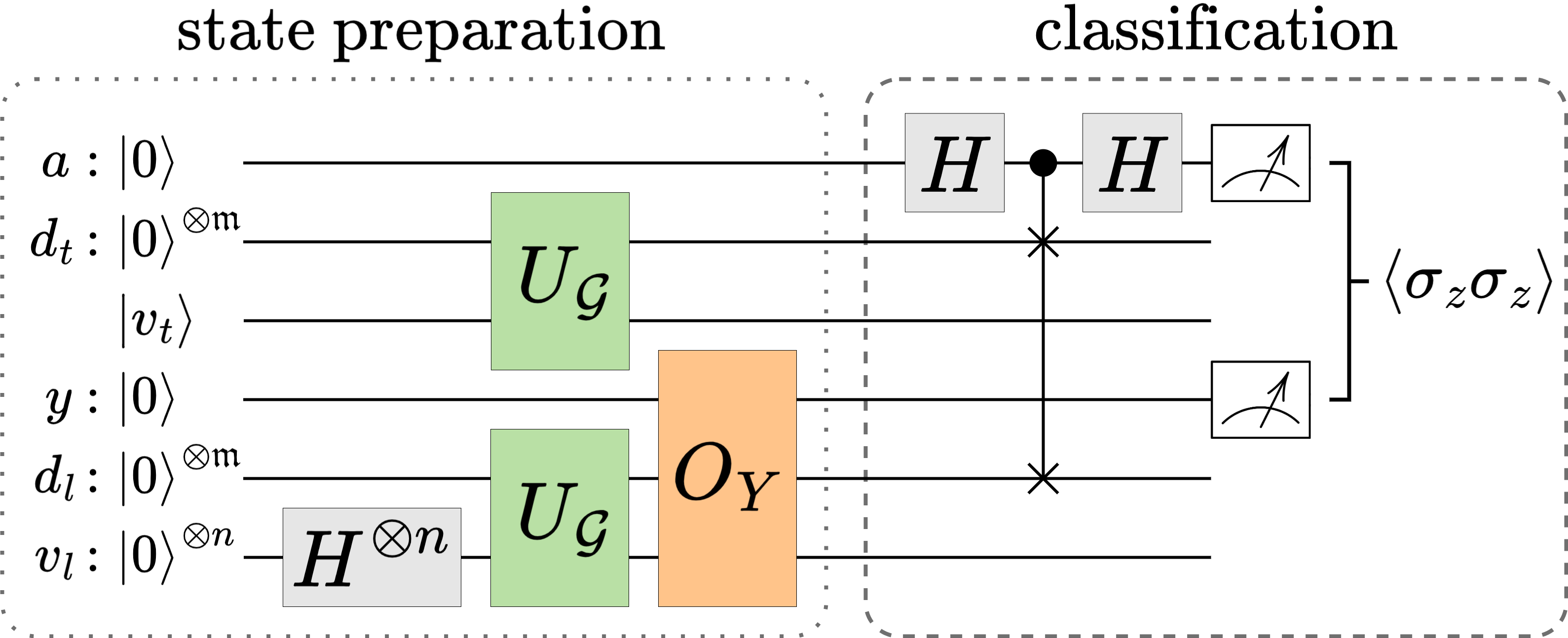}    
    \caption{\label{fig:qmme+swap-test}
    \textbf{Full cancer gene prediction pipeline with the QMME protocol as part of state preparation.} 
    The first register corresponds to the ancilla qubit \(a\), the second to the test data qubits \(d_t\), the third to the index qubits of the test node \(v_t\), the fourth stores the label qubit \(y\), the fifth the labeled data qubits \(d_l\), and the sixth the index qubits of the labeled nodes \(v_l\). The operator \(U_\mathcal G\), together with the Hadamard gate and the oracle \(O_Y\), prepares the quantum state, which is then processed by the classifier $\mathcal{C}$, i.e., the swap-test classifier. Node labels \(\hat{y}_v \in \{0,1\}\) are predicted via a swap-test and two-qubit measurement statistics on the ancilla and label registers. An outcome of label one indicates that the test gene is predicted as a cancer driver.} 
\end{figure*}
\noindent We define the QMME unitary operator \(U_{\mathcal{G}}\) as
\begin{equation}
U_\mathcal{G} : \ket{v} \ket{0}^{\otimes \mathfrak{m}} \mapsto \ket{v} \ket{\boldsymbol{\psi}_v},
\label{eq:qmme-unitary}
\end{equation}
and refer to this mapping as the \textit{quantum node embedding}. The second register, consisting of \(\mathfrak{m}\) qubits, is called the \textit{data register}. The total number of qubits is \(\mathfrak{m} = 2\log N + 8\) (excluding work registers), and generally scales as \(\mathcal{O}(\log N)\). 
\noindent A schematic of the operator \(U_\mathcal{G}\) circuit is shown in~\Cref{fig:qmme-schematics}, and step-by-step derivation of the quantum node embedding in~\cref{eq:qmme-unitary} is presented in the Appendix.
The resulting state \(\ket{\boldsymbol{\psi}_v}\) consists of two components:
\begin{equation}
\ket{\boldsymbol{\psi}_v} = \ket{\boldsymbol{\psi}_m} + \ket{\boldsymbol{\psi}_\perp},
\end{equation}
with \(\ket{\boldsymbol{\psi}_m}\) denoting the \textit{neighborhood-moment component}, which captures the quantum analogue of the classical moment-feature embedding \(\boldsymbol{m}_v \in \mathbb{R}^8\) defined in~\cref{eq:origin-moment-embedding-function}, and \(\ket{\boldsymbol{\psi}_\perp}\) corresponding to an orthogonal component.

According to the circuit in~\Cref{fig:qmme-schematics}, \(\ket{\boldsymbol{\psi}_m}\) is defined as
\begin{align}
\ket{\boldsymbol{\psi}_m} 
&= \ket{\boldsymbol{c} \odot \boldsymbol{m}_v} \ket{0}^{\otimes (\mathfrak{m} - 3)} \label{eq:psi_m}\\ 
&= \sum_{c \in [2]} \frac{1}{2^{3/2} \, s^c} \sum_{i \in [4]} \varphi_i(\mathcal{X}_v^c) \ket{c, i} \ket{0}^{\otimes (\mathfrak{m} - 3)} \label{eq:psi_m_detail}
\\
&= \sum_{j\in [8]} c_j \cdot m_{v,j} \ket{j} \ket{0}^{\otimes (\mathfrak{m} - 3)},
\end{align}
where \(\odot\) denotes the element-wise (Hadamard) product, \(\varphi_i(\mathcal{X}_v^c)\) are the \(i\)th-order moment of the neighborhood \(\mathcal{X}_v^c\), and the constant vector \(\boldsymbol{c} \in \mathbb{R}^8\) is
\begin{equation}
\boldsymbol{c} = \frac{1}{2^{3/2}} \left[\frac{1}{s}, \, \frac{1}{s}, \, \frac{1}{s}, \, \frac{1}{s}, \, \frac{1}{s^2}, \, \frac{1}{s^2}, \, \frac{1}{s^2}, \, \frac{1}{s^2} \right].
\label{eq:c-constant-vector}
\end{equation}
\noindent The orthogonal component \(\ket{\boldsymbol{\psi}_\perp}\) of the resulting state is
\begin{equation}
\ket{\boldsymbol{\psi}_\perp} = \sum_{a \neq 0} \ket{\boldsymbol{\psi}_{\perp,a}} \ket{a},
\end{equation}
with \(\ket{a}\) representing computational basis states in the \((\mathfrak m-3)\) data qubits, and \(\ket{\boldsymbol{\psi}_{\perp,a}} \in \mathbb{R}^{2^{(\mathfrak{m}-3)}}\).

As an example, take the mean of the mutation scores of the immediate neighbors of gene \(v\), i.e., \(m_{v,1} = \varphi_1(\mathcal{X}_v^1)\). As in~\cref{eq:psi_m_detail}, the amplitude of the state \(\ket{0_c, 0_i, 0^{ (\mathfrak{m} - 3)}}\) is given by \(\frac{\varphi_1(\mathcal{X}_v^1)}{2^{3/2} s}\). This amplitude is obtained through the following sequence of operations, illustrated in~\Cref{fig:qmme-schematics}:
\begin{enumerate}
    \item preparing an equal superposition over register \(\ell_1\);
    \item \(O_L\) retrieves the immediate neighbors' indices;
    \item operator \(O_X\) encodes the neighbors' feature values into the \(\ket{0}\) component of the first ancilla qubit;
    \item uncomputing the basis register \(\ell_1\) via \(O_L^T\) leaves the \(\ket{0}^{\otimes n}\) component amplitude proportional to the sum of neighbors' features, while others form interference patterns with alternating signs;
    \item finally, the ancilla rotation \(O_{K^{-1}}\) normalizes by the degree \(k_v^c\), resulting in the mean of \(\mathcal X_v^c\) encoded in the amplitude of the rotated ancillas at zero.
\end{enumerate}
\noindent 
The sequence above describes the action of the unitary \(U_\mathcal G\) for \(c,i=0\). Extending the same steps to all \(c \in [2]\) and \(i \in [4]\) covers both immediate and second neighbors and encodes neighborhood moments of orders one through four. In general, applying \(U_\mathcal G\) yields
\begin{align}
\bra{v,c,i,0^{(\mathfrak{m}-3)}} U_{\mathcal{G}} \ket{v,0^{\mathfrak{m}}} = \frac{1}{2^{3/2} s^{c}} \sum_{x_u \in \mathcal X_v^c}\frac{x_u^i}{k_v^c}=\frac{\varphi_i(\mathcal{X}_v^c)}{2^{3/2} s^c}.
\end{align}

\section{Quantum Classification Framework}
\noindent
The quantum node embedding operator \(U_{\mathcal{G}}\) maps each node in a feature-labeled network to its statistical moments, amplitude-encoded within a quantum state \(\ket{\boldsymbol \psi_v}\). Given a protein-protein interaction (PPI) network as input, the QMME protocol encodes local statistical moments of the MutSig \(P\)-values. Depicted in~\Cref{fig:qmme+swap-test}, the quantum classification framework assumes query access to the class labels \(y_{v_l}\in\{0,1\}\) for each labeled node \(v_l\in V_l\):
\begin{equation*}
    \textbf{label query: } \text{given } v_l \in V_l, \text{ returns the label } y_{v_l} {,}
\end{equation*}
\noindent with a quantum equivalent defined by the operator \(O_Y\):
\begin{align}
    O_Y &:\ket{v_l} \ket{0} \mapsto \ket{v_l}\ket{y_{v_l}},
    \label{eq:label-oracle}
\end{align}
\noindent encoding the class label \(y_{v_l}\in\{0,1\}\) in a single qubit. 
The proposed quantum classification framework proceeds as follows: after preparing the equal superposition over labeled nodes, the embedding unitary \(U_\mathcal G\) is applied both to this register \(v_l\) and to the test node register \(v_t\). The class labels are loaded onto the label qubit via \(O_Y\). Classification follows via the swap-test classifier~\cite{blank2020quantum}, where the measurement statistics determine the final class prediction of the test node. 
The full pipeline may be described as a sequence of quantum state evolutions:
\begin{equation}
\begin{aligned}
    &\ket{0} \ket{0}^{\otimes \mathfrak{m}} \ket{v_t} \ket{0} \ket{0}^{\otimes \mathfrak{m}} \ket{0}^{\otimes n} \\
    &\xrightarrow{H^{\otimes n}_{v_l}} 
    \frac{1}{\sqrt{|V_l|}} \sum_{v_l \in V_l} \ket{0} \ket{0}^{\otimes \mathfrak{m}} \ket{v_t} \ket{0} \ket{0}^{\otimes \mathfrak{m}} \ket{v_l} \\
    &\xrightarrow{U_\mathcal{G}}
    \frac{1}{\sqrt{|V_l|}} \sum_{v_l \in V_l} \ket{0} \ket{\boldsymbol{\psi}_t} \ket{v_t} \ket{0} \ket{\boldsymbol{\psi}_l} \ket{v_l} \\
    &\xrightarrow{(O_Y)_{v_l,y}}
    \frac{1}{\sqrt{|V_l|}} \sum_{v_l \in V_l} \ket{0} \ket{\boldsymbol{\psi}_t} \ket{v_t} \ket{y_l} \ket{\boldsymbol{\psi}_l} \ket{v_l} \\
    &\xrightarrow{H_a \cdot \text{c-swap} \cdot H_a}\ket{\boldsymbol \psi_f}=
    \frac{1}{2\sqrt{|V_l|}} \sum_{v_l \in V_l} \left( \ket{0} \ket{\boldsymbol{\psi}_+} + \ket{1} \ket{\boldsymbol{\psi}_-} \right) \ket{v_l},
\end{aligned}
\end{equation}
\noindent where \(\ket{\boldsymbol \psi_{\pm}}=\ket{\boldsymbol \psi_t}(\ket{v_t}\ket {y_l} )\ket{\boldsymbol \psi_l}\pm\ket{\boldsymbol \psi_l}(\ket{v_t}\ket {y_l} )\ket{\boldsymbol \psi_t}\). 
\begin{figure*}[!t]
    \includegraphics[width=\textwidth]{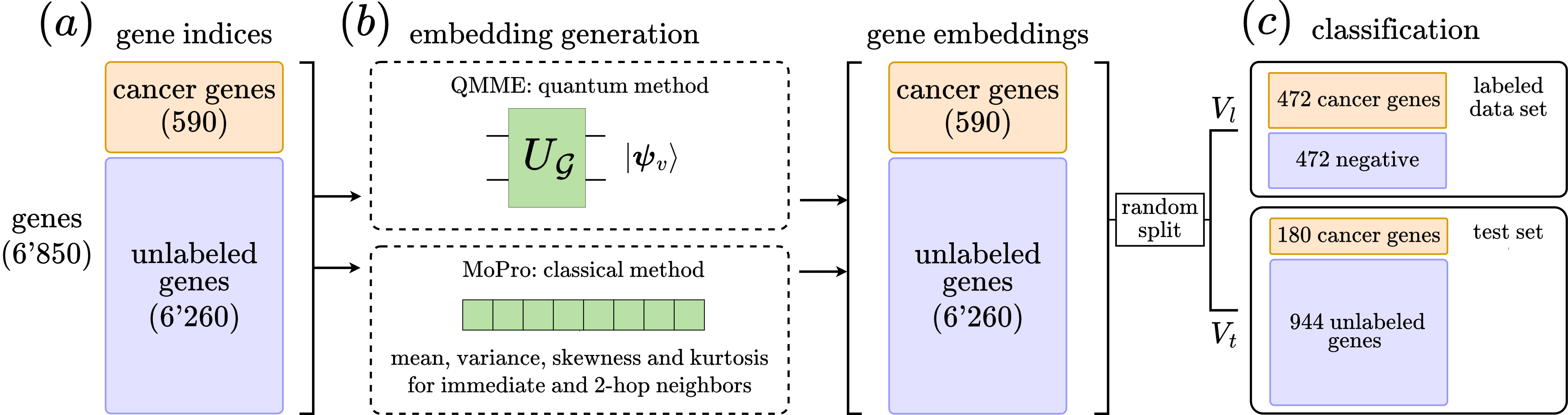}    
    \caption{\label{fig:compare-methods}
    \textbf{Experimental setup for QMME and classical embedding comparison and performance evaluation.}
    \textbf{(a)} The dataset consists of 6 850 genes, of which 590 (in orange) are known cancer driver genes~\cite{10.1093/nar/gky1015}. The remaining 6 260 genes have unknown labels, as they are not part of the CGC dataset.
    \textbf{(b)} Embeddings for each gene \(v\) are obtained via two methods: the Quantum Multi-order Moment Embedding (QMME) protocol, and the Moment Propagation (MoPro) embedding with hyperparameter \(t=1\)~\cite{gumpinger2020prediction}. The QMME approach is implemented via algebraic computations on classical software.
    \textbf{(c)} In \textit{each random split}, 80\% of the positive genes are selected and combined with an equal number of unlabeled genes (assigned a negative class label) to form the labeled training set $V_l$. The remaining genes form the test set $V_t$, whose labels are used only for evaluation. For both embedding methods, kernel matrices are computed and the expectation values in~\Cref{eq:expectation-values} are evaluated on all test genes. This process is repeated across 50 random splits, and performance metrics are averaged.
    }  
\end{figure*}
The swap-test classifier computes the fidelity, i.e., the squared overlap between two states. This yields \( |\langle \boldsymbol \psi_t | \boldsymbol \psi_l \rangle|^2 \) for the two pure states \( |\boldsymbol \psi_t\rangle \) and \( |\boldsymbol \psi_l\rangle \).
Fidelity serves as a similarity measure (one if the states are identical, zero if they are orthogonal), making it a natural distance measure in the quantum feature space. Classification thus relies directly on distances between training and test data, unifying feature mapping and classification into a single quantum procedure without requiring an intermediate classical support vector machine.

The measurement and post-processing involves evaluating the expectation value of the two-qubit observable \(\sigma_z^{(a)} \sigma_z^{(l)}\), where the superscripts indicate the ancilla and label qubit. For the final state \(\ket{\boldsymbol \psi_f}\), \(\langle \sigma_z^{(a)} \sigma_z^{(l)} \rangle\) is given by
\begin{align}
\langle \sigma_z^{(a)} \sigma_z^{(l)} \rangle &= \mathrm{Tr} \left( \sigma_z^{(a)} \sigma_z^{(l)} \ket{\boldsymbol \psi_f}\bra{\boldsymbol \psi_f} \right) \\
&= \frac{1}{|V_l|} \sum_{v_l \in V_l} (-1)^{y_{v_l}} |\langle \boldsymbol \psi_t |\boldsymbol \psi_{l}\rangle|^2.
\label{eq:expectation-values}
\end{align}
\noindent In practice, these statistics are obtained by repeating \(r\) times the two-qubit projective measurement in the \(\sigma_z\) basis. Using \(\text{Tr}(\sigma_z\ket{y_{v_l}}\bra{y_{v_l}})=1\) for \(y_{v_l}=0\) and \(-1\) for \(y_{v_l}=1\), the expectation value is calculated as 
\begin{equation}
    \langle \sigma_z^{(a)} \sigma_z^{(l)} \rangle = \frac{1}{r} \left( c_{00} - c_{01} - c_{10} + c_{11} \right),
    \label{eq:expectat-val-counts}
\end{equation}
where \(c_{a l}\) counts measurements with the ancilla in state \(a\) and the label qubit in state \(l\). 
The test data are classified as 0 if \(\langle \sigma_z^{(a)} \sigma_z^{(l)} \rangle>0\) and 1 if negative, i.e.,
\begin{equation}\label{eq:test-labels-predict}
\begin{aligned}
\hat{y}_v &= \frac{1}{2} \left( 1 - \mathrm{sgn} \, \langle \sigma_z^{(a)} \sigma_z^{(l)} \rangle \right) \\
&= \mathcal{C}\left(U_\mathcal{G}\ket{v} \ket{0}^{\otimes{\mathfrak m}} \right)
\end{aligned}
\end{equation}
which resembles the earlier formulation in~\Cref{eq:goal}. Here, \(\mathcal C\) is the swap-test classifier and \(U_\mathcal G\) corresponds to the Quantum Multi-order Moment Embedding (QMME), encoding local properties in the network \(\mathcal G\).
\section{Results and Analysis}
\noindent  To test our quantum classification approach, we applied it to the OmniPath protein-protein interaction (PPI) network~\cite{turei2016omnipath, turei2021integrated}, accessed via the \texttt{OmnipathR} R package. 
OmniPath integrates data from over a hundred molecular biology resources, capturing both direct physical binding, such as proteins forming complexes or modifying each other enzymatically, and functional associations, such as participation within the same metabolic pathway or biological process. 
This network is widely used in computational biology~\cite{carvalho2019open, menden2019community, zhou2019metascape, browaeys2020nichenet, dou2020proteogenomic}, and offers a framework for studying gene interactions. Nodes represent genes, and edges correspond to protein-level interactions.
We extracted all pairwise gene interactions, filtered out entries with missing gene symbols, and removed duplicate edges, resulting in a network of 79 940 interactions. 
Each gene (node) is annotated with its \(-\log_{10}\) MutSig \(P\)-value~\cite{lawrence2014discovery}, capturing the statistical significance of mutations in cancer. Genes without MutSig values were removed from the network.
The resulting network contains 27 075 interactions and 6 850 genes, forming the gene set for node classification. 
Positive ground-truth labels are obtained from the Cancer Gene Census (CGC) in the COSMIC database~\cite{10.1093/nar/gky1015}. Tier 1 genes have documented cancer-relevant activity and mutations that promote oncogenic transformation, while Tier 2 genes show strong but less extensive evidence.
Both Tier 1 and Tier 2 CGC genes are treated as positives in our experimental setup, totaling 590 known cancer genes (\(\sim 8.6\%\) of the dataset). The remaining genes are left unlabeled. 
\noindent In the constructed dataset, the number of positive samples (cancer driver genes) is much smaller than the pool of negative or unlabeled samples. If left uncorrected, this class imbalance biases the classifier toward predicting negatives. To mitigate this, the labeled training set \(V_l\) is constructed for each random split by selecting \(80\%\) of the known cancer driver genes and pairing them with an equal number of randomly sampled unlabeled genes, which are provisionally assigned to the negative class (Fig.~\ref{fig:compare-methods}c). This corresponds to a balanced set of 944 genes (472 positives and 472 negatives). Although some of the unlabeled genes may be undiscovered drivers, treating them as negatives provides a balanced ground truth to inform the classifier. The remaining genes form the disjoint test set \(V_t\), and this process is repeated across 50 independent splits, with performance metrics averaged over all splits.  
Balancing is especially important in the quantum setting, as the first step of state preparation constructs a superposition over labeled nodes, and an imbalanced distribution would induce skewed amplitudes. 
\noindent For both QMME and classical Moment Propagation (MoPro) embeddings, we construct a kernel matrix \(K\) from the embedding vectors and use the same distance-based classification procedure to predict labels. This ensures performance differences are attributable solely to the embeddings rather than classifier choice, providing a fair comparison. Classification labels are predicted from expectation values as in eq.~\eqref{eq:test-labels-predict}. Statistical significance of the differences between QMME and MoPro metrics is assessed via paired sample t-tests.  

\begin{table}[h]
    \centering
    \caption{\textbf{Cancer gene classification performance (\%).} Comparison of Quantum Multi-order Moment Embeddings (QMME) and Moment Propagation (MoPro) embeddings over 50 random splits. All metric differences are statistically significant except AUPRC.}
    \resizebox{1.02\columnwidth}{!}{%
    \begin{tabular}{l@{\hskip 0.5em}c@{\hskip 0.5em}c@{\hskip 0.05em}c}
        \toprule
        \textbf{Metric} & \textbf{QMME} & \textbf{MoPro} & \textbf{t-test} \\
        \midrule
        Accuracy  & 37.2 & 53.1 & $t=-6.829, p=0.00000122$ \\
        Precision & 2.1  & 2.3  & $t=-3.486, p=0.00105$ \\
        Recall    & 65.9 & 53.3 & $t=5.341, p=0.000238$ \\
        F1-Score  & 4.0  & 4.4  & $t=-3.432, p=0.00123$ \\
        AUPRC     & 1.9 \!$\pm$\! 0.2 & 1.8\! $\pm$\! 0.2 & --- \\
        \bottomrule
    \end{tabular}%
    }
    \label{tab:results}
\end{table}

\noindent Compared with the original MoPro pipeline~\cite{gumpinger2020prediction}, which applies multiple classifiers with optimized hyperparameters (logistic regression, random forest, SVM, gradient boosting) and propagation steps \(t>1\), ours uses only one-step moment embeddings and the same distance-based classifier for both QMME and MoPro. This avoids unfair advantages from classifier choice or propagation, focusing solely on the quality of the embedding representations. In general, both classical and quantum approaches yield comparable outcomes, as their underlying principles do not differ substantially.
Nevertheless, in detail, while absolute AUPRC values in~\Cref{tab:results} are low due to the rarity of positive samples in the test set, QMME embeddings achieve a slightly higher mean AUPRC (0.0191) compared with MoPro (0.0177), and also exhibit lower variance across splits (0.0015 vs 0.0016).
\noindent This indicates that QMME provides a more consistent ranking of true positives across random splits. The QMME embeddings achieve substantially higher recall (65.9\%) compared with MoPro (53.3\%), corresponding to a 12.6 percentage point improvement. This indicates that QMME is more effective at identifying true cancer driver genes (and rare positives) despite the low precision and overall accuracy.

While the framework is described using quantum states and operations for conceptual alignment with quantum computing principles, all quantum procedures, including the preparation of embeddings via $U_\mathcal{G}$, the application of the swap-test, and the evaluation of expectation values, are implemented as classical simulations. Specifically, the embedding vectors and their inner products are computed algebraically on a classical computer using parallelized C++ code, without executing any operations on physical quantum hardware. This approach allows practical evaluation of the protocol while preserving the theoretical structure of the QMME and swap-test framework.
\section{Complexity Analysis}
\label{sec:complexity}
\noindent
We analyze the complexity of the cancer driver gene prediction pipeline in~\Cref{tab:complexity}. 
Let \( d' \) denote the precision (in qubits) used to encode both feature values and node degrees. 
The number of oracle calls required for a single application of \(U_\mathcal G\) is constant. 
The Hadamard gates contribute a cost that scales as \(\mathcal{O}(\log s)\) within the unitary operator \(U_\mathcal G\), and as \(\mathcal{O}(\log |V_l|)\) across the entire quantum pipeline. 
Overall, the dominant contribution to the total gate complexity arises from arithmetic operations that are part of oracle subroutines (see Appendix~\ref{appendix:complexity-analysis-in-detail} for a detailed breakdown).

\begin{table}[h]
    \centering
    \caption{\textbf{Complexity analysis of the full cancer driver gene prediction pipeline}, including state preparation (with the QMME protocol) and classification. 
    Reported in terms of gate count, qubit requirements, circuit depth, and query complexity (calls to primitive oracles). 
    }
    \label{tab:complexity}
    \begin{tabular}{lcc}
        \toprule
        \textbf{Metric} & \textbf{Complexity} \\
        \midrule
        Gate complexity   & \(\mathcal O(\text{polylog}\, N + d')\) \\
        Qubit count       & \(\mathcal O(\log\, N + d')\) \\
        Circuit depth     & \(\mathcal O(\log N)\) \\
        \midrule
        Query complexity  & \(\mathcal O(\log^2 s)\) \\
        \bottomrule
    \end{tabular}
\end{table}
\noindent Circuit depth is the number of sequential gate layers, i.e., the computation length, and measures time complexity. The number of qubits is the size of the quantum register, which measures space (memory) complexity. For the embedding circuit \(U_\mathcal G\), time complexity is independent of the network size, and space complexity scales polylogarithmically. Considering the entire classification pipeline, the swap-test classifier, which relies on controlled-swap operations to estimate quantum state overlap, introduces a logarithmic dependence on the feature dimension \(2^{\mathfrak m}\) in its circuit depth. Since \(\mathfrak m = \mathcal O(\log N)\), this results in an overall logarithmic dependence on the network size \(N\).
\section{Discussion and Conclusions}
\noindent
In this work, we introduce an almost fully quantum pipeline for cancer driver gene prediction based on a swap-test classifier, requiring no classical optimization loops. All stages of encoding, embedding, and inference are performed within the quantum domain, requiring only minimal data access assumptions and limited classical post-processing. Our approach demonstrates polylogarithmic resource scaling together with competitive performance on protein-protein interaction (PPI) network data.
The proposed \textit{Quantum Multi-order Moment Embedding} (QMME) is a modular, extensible and general-purpose architecture for encoding local graph statistics into quantum states. While we focus on binary classification of genes as cancer drivers using PPI networks (with nodes representing genes and MutSig \(P\)-values as features), QMME is broadly applicable to node-level classification tasks on feature-labeled graphs across domains~\cite{bi2023mm}.
Additional feature preprocessing or more sophisticated classification algorithms can be incorporated with minimal modification, allowing easy adaptation to different network types or learning objectives.
Future directions include extending QMME to capture higher-order moments, supporting weighted and multi-layered graphs, enabling multi-class classification, and implementing resource-efficient variants suitable for near-term quantum hardware.

\vspace{2em}
\noindent \textbf{Acknowledgements}

\begin{scriptsize}
\noindent 
This research work has been supported by FCT/MECI through project QuNetMed (Grant No. 2022.05558.PTDC, DOI: 10.54499/2022.05558.PTDC), 
and through Grant UIDB/50008/2020 (DOI: 10.54499/UIDB/50008/2020); Instituto de Telecomunicações. We also acknowledge the open-source community and resources that made this research possible.
\end{scriptsize}

\vspace{2em}
\noindent \textbf{Author Contributions}

\begin{scriptsize}
\noindent BC and DM conceived the initial concept of the project. AW made substantial contributions to the machine learning components. DM assisted with the complexity analysis of the quantum circuit.
PM developed the framework, performed all calculations and simulations, and developed the main tools, in close collaboration with BC and AW. 
PM drafted the manuscript, and BC, AW and DM provided critical review and revisions.
\end{scriptsize}

\vspace{2em}
\noindent \textbf{Data and Code Availability}

\begin{scriptsize}
\noindent The implementation and all data used in this work are openly available at 
\href{https://github.com/apatriciamarques/Network-Based-Prediction}{https://github.com/apatriciamarques/Network-Based-Prediction}.
\end{scriptsize}

\vspace{2em}
\noindent \textbf{Contact}

\begin{scriptsize}
\noindent Correspondence: \texttt{patriciasmarques@tecnico.ulisboa.pt}.
\end{scriptsize}

\vspace{2em}
\noindent \textbf{Competing Interests}

\begin{scriptsize}
\noindent The authors declare no competing interests.
\end{scriptsize}

\nocite{*}

\bibliographystyle{quantum}
\bibliography{bibfile}

\normalsize
\clearpage
\onecolumngrid
\appendix
\allowdisplaybreaks
\section{Quantum Node Embedding Step-by-step Derivation}
\label{appendix:mapping}
\noindent 
Let \(U_r\) denote the operator \(U\) acting on register \(r\), with identity on all other registers. The notation \(\mathrm{c}\text{-}U_r\) represents the controlled application of \(U_r\), with the control qubit in register \(c\) in state \(\ket{1}\). Similarly, \(\mathrm{c}^{(0)}\text{-}U_r\) denotes control on the \(\ket{0}\) state of \(c\).  
Below, we describe the evolution of the quantum state through the Quantum Multi-order Moment Embedding (QMME) protocol defined in~\Cref{eq:qmme-unitary}, step by step, including a brief description and gate complexity estimate for each transformation (e.g., oracle applications and other unitary operations).
\newcommand{\zero}[1]{0^{#1}}
\newcommand{\ketzero}[1]{\ket{0^{#1}}}
\newcommand{\rvl}{r(v,\ell_1)}
\newcommand{\rrvll}{r(\mathcal{N}^1_v(\ell_1), \ell_2)}
\newcommand{\garbage}{\ket{\perp}}
\small
\begin{align*}
&\text{Start with the initial state. The node \(v\) is encoded, and all other qubits are initialized to zeros.}
\\
&\quad\ket{v,0^{ \mathfrak m}} \quad \text{where } \mathfrak m = 2\log N +8, \text{omitting work qubits.}
\\
&\text{Put the register \(l_1\) into an equal superposition over all possible neighbor indices \(\ell_1 \in [s]\); counts as \(\mathcal O(\log s)\)  in gate complexity.}
\\
&\quad\mathrel{\xrightarrow{I_{n-\mathfrak{s}}\otimes H^{\otimes \mathfrak{s}}_{l_1}}} 
\frac{1}{\sqrt{s}} \ket{v,0^{{3}}} \sum_{\ell_1\in[s]} \ket{\ell_1, \zero{(n + 5)}} 
\\
&\text{Map each \(\ell_1\) in superposition to the actual index of the corresponding neighbor of \(v\); requires \(\mathcal O(\text{polylog}\, N)\) elementary gates.}
\\
&\quad\mathrel{\xrightarrow{(O_L)_{v,l_1}}} 
\frac{1}{\sqrt{s}} \ket{v,0^{{3}}} \sum_{\ell_1\in[s]} \ket{\rvl, \zero{(n+5)}} 
\\
&\text{Put the control qubit \(c\) into superposition (for immediate and second-order neighborhoods); requires a single elementary gate.}
\\
&\quad\mathrel{\xrightarrow{H_c}} 
\frac{1}{\sqrt{2s}} \ket{v} \sum_{c\in[2]} \sum_{\ell_1\in[s]} \ket{c, 0^{{2}}, \rvl, \zero{(n + 5)}} 
\\
&\text{If \(c=1\), create a superposition over all second-neighbor indices \(\ell_2\), capturing the two-hop neighborhood (\(\mathcal O(\log s)\) gates).}
\\
&\quad\mathrel{\xrightarrow{I_{n-\mathfrak{s}}\otimes \mathrm{c}\text{-}H^{\otimes \mathfrak{s}}_{\ell_2}}} 
\frac{1}{\sqrt{2s}} \ket{v} \sum_{\ell_1\in[s]} \Biggl( \ket{0_c, 0^{{2}}, \rvl, \zero{(n+5)}} 
+ \frac{1}{\sqrt{s}} \sum_{\ell_2\in[s]} \ket{1_c, 0^{{2}}, \rvl, \ell_2,\zero{5}} \Biggr)
\\
&\text{If \(c=1\), retrieve the indexes of second-order neighbors of \(v\); requires \(\mathcal O(\text{polylog}\, N)\) gates.}
\\
&\quad\mathrel{\xrightarrow{\mathrm{c}\text{-}(O_L)_{\ell_1,\ell_2}}} 
\frac{1}{\sqrt{2s}} \ket{v} \sum_{\ell_1\in[s]} \Biggl( \ket{0_c, 0^{{2}}, \rvl, \zero{(n+5)}} 
+ \frac{1}{\sqrt{s}} \sum_{\ell_2\in[s]} \ket{1_c, 0^{{2}}, \rvl, \rrvll,\zero{5}} \Biggr)
\\
&\text{Create an equal superposition over the four moment indices \(i\); requires \(\mathcal O(1)\) gates.}
\\
&\quad\mathrel{\xrightarrow{H^{\otimes 2}_i}} 
\frac{1}{2^{3/2}\sqrt{s}} \ket{v} \sum_{\ell_1\in[s]} \sum_{i\in[4]} \Biggl( \ket{0_c, i, \rvl, \zero{(n+5)}} 
+ \frac{1}{\sqrt{s}} \sum_{\ell_2\in[s]} \ket{1_c, i, \rvl, \rrvll,\zero{5}} \Biggr)
\\
&\text{If \(c=0\), amplitude-encode the \(i\)-th power of the immediate neighbors’ features; gate cost \(\mathcal O(\text{polylog}\, N + d')\).}
\\
&\quad\mathrel{\xrightarrow{\mathrm{c}^{(0)}\text{-}(O_X)_{i,a,l_1}}} 
\frac{1}{2^{3/2}\sqrt{s}} \ket{v} \sum_{\ell_1,i} \Biggl( \bigl(x_{\rvl}^i\ket{0_c, i, \rvl, \zero{(n+4)}}+\ket{\perp}\bigr)\ket 0 
+ \frac{1}{\sqrt{s}} \sum_{\ell_2\in[s]} \ket{1_c, i, \rvl, \rrvll, \zero{5}} \Biggr)
\\
&\text{If \(c=1\), amplitude-encode the \(i\)-th power of the second-hop neighbors’ features; \(\mathcal O(\text{polylog}\, N + d')\) in gate complexity.}
\\
&\quad\mathrel{\xrightarrow{\mathrm{c}\text{-}(O_X)_{i,a,\ell_2}}} 
\frac{\ket{v}}{2^{3/2}\sqrt{s}} \sum_{\ell_1,i} \Biggl( x_{\rvl}^i \ket{0_c, i, \rvl, \zero{(n+5)}} 
+ \frac{1}{\sqrt{s}} \sum_{\ell_2\in[s]} x_{\rrvll}^i \ket{1_c, i, \rvl, \rrvll,\zero{5}} \Biggr)
+ \ket{v} \garbage \ket{0} 
\\
&\text{If \(c=1\), rewrite the second-hop neighbor label \(\rrvll\) back into its index \(\ell_2\); gate cost \(\mathcal O(\text{polylog}\, N)\).}
\\
&\quad\mathrel{\xrightarrow{\mathrm{c}\text{-}(O_L^T)_{\ell_1,\ell_2}}} 
\frac{1}{2^{3/2}\sqrt{s}} \ket{v} \sum_{\ell_1,i} \Biggl( x_{\rvl}^i \ket{0_c, i, \rvl, \zero{(n+5)}} 
+ \frac{1}{\sqrt{s}} \sum_{\ell_2\in[s]} x_{\rrvll}^i \ket{1_c, i, \rvl, \ell_2,\zero{5}} \Biggr)
+ \ket{v} \garbage \ket{0} 
\\
&\text{If \(c=1\), uncompute the \(\ell_2\) superposition, resetting the index register (sum second-neighbors' features); requires \(\mathcal O(\log s)\) gates.} 
\\
&\quad\mathrel{\xrightarrow{I_{n-\mathfrak{s}}\otimes \mathrm{c}\text{-}H^{\otimes \mathfrak{s}}_{\ell_2}}} 
\frac{1}{2^{3/2}\sqrt{s}} \ket{v} \sum_{\ell_1,i} \Biggl( x_{\rvl}^i \ket{0_c, i, \rvl} 
+ \frac{1}{s} \sum_{x' \in \mathcal{X}_{\rvl}^1} x'^i \ket{1_c, i, \rvl} \Biggr) \ketzero{(n + 5)} 
+ \ket{v} \garbage \ket{0} 
\\
&\text{Rewrite the immediate neighbor lookup \(\rvl\) back to its index \(\ell_1\); requires \(\mathcal O(\text{polylog}\, N)\) gates.}
\\
&\quad\mathrel{\xrightarrow{(O_L^T)_{v,l_1}}}  
\frac{1}{2^{3/2}\sqrt{s}} \ket{v} \sum_{\ell_1,i} \Biggl( x_{\rvl}^i \ket{0_c, i, \ell_1} 
+ \frac{1}{s} \sum_{x' \in \mathcal{X}_{\rvl}^1} x'^i \ket{1_c, i, \ell_1} \Biggr) \ketzero{(n+5)} 
+ \ket{v} \garbage \ket{0} 
\\
&\text{Uncompute the \(\ell_1\) superposition, resetting the index register (sum immediate neighbors' features); requires \(\mathcal O(\log s)\) gates.} 
\\
&\quad\mathrel{\xrightarrow{I_{n-\mathfrak{s}}\otimes H^{\otimes \mathfrak{s}}_{l_1}}}  
\frac{1}{2^{3/2}s} \ket{v} \sum_{i\in[4]} \Biggl( \sum_{x' \in \mathcal{X}_{\rvl}^1} x'^i \ket{0_c} 
+ \frac{1}{s} \sum_{x'' \in \mathcal{X}_v^2} x''^i \ket{1_c} \Biggr) \ket{i, \zero{(2n + 5)}} 
+ \ket{v} \garbage \ket{0} 
\\
&\text{Normalize by \(v\)'s degree, amplitudes are averaged neighbor statistics rather than sums; gate complexity \(\mathcal O(\text{polylog}\, N + d')\).}
\\
&\quad\mathrel{\xrightarrow{(O_{K^{-1}})_{v,k}}}  
\frac{1}{2^{3/2}s} \ket{v} \sum_{i\in[4]} \Biggl( \sum_{x' \in \mathcal{X}_{\rvl}^1} x'^i \ket{0_c} 
+ \frac{1}{s} \sum_{x'' \in \mathcal{X}_v^2} x''^i \ket{1_c} \Biggr) \frac{1}{k_v^c} \ket{i, \zero{(2n + 5)}} 
+ \ket{v} \garbage
\\ 
&\quad= \ket{v} \Biggl( \sum_{i,c} \frac{1}{2^{3/2} s^{c}} \sum_{x_u \in \mathcal X_v^c}\frac{x_u^i}{k_v^c}\ket{c,i,0^{ ({2n + 5})}} + \garbage \Biggr) 
\\ 
&\quad= \ket{v} \Biggl( \sum_{i,c} \frac{1}{2^{3/2} s^{c}} \varphi(\mathcal{X}^c_v) \ket{c,i,0^{ ({2n + 5})}} + \garbage \Biggr) 
\\
&\quad= \ket{v} \Biggl( \sum_{j\in [8]} c_j m_{v,j} \ket{j,0^{({2n + 5})}} + \garbage \Biggr) \quad \text{with \(\boldsymbol c\) defined as in~\cref{eq:c-constant-vector}.}
\end{align*}

\normalsize
\section{Complexity Analysis in Detail}
\label{appendix:complexity-analysis-in-detail}
\noindent
We can construct a quantum circuit to implement the primitive oracles \(O_L\) and \(\tilde O_X\) with, respectively, \(\mathcal O(\mathrm{polylog}\,N)\) and \(\mathcal O(\mathrm{polylog}\,N + d')\) elementary gates.  
\(O_L\) performs a neighbor lookup, while \(\tilde O_X\) encodes the \(d'\)-bit feature vector of a node into a binary register. Thus, the cost of implementing neighborhood and feature access oracles is polylogarithmic in the graph size, with an additional linear factor in the precision \(d'\) for \(\tilde O_X\). Controlled versions of these oracles, as well as their inverses, inherit the same asymptotic cost.
We assume costs consistent with standard data structures for this problem class, as in Remark 3 from Tong \textit{et al.}~\cite{tong2021fast}:
\begin{quote}
If we consider a diagonal matrix \(D \in \mathbb{R}^{N\times N}\) and assume that each diagonal entry is accessible via an oracle
\[
O_D\ket{i}\ket{0^l} = \ket{i}\ket{D_{ii}}, \quad i \in [N],
\]
where \(\ket{D_{ii}}\) is the \(l\)-bit binary representation of \(D_{ii}\), then each diagonal entry of \(D\) can be efficiently computed using a classical Boolean circuit of size \(\mathcal O(\mathrm{polylog}\,N)\) with \(m=\mathcal O(\mathrm{polylog}\,N)\) ancilla bits. In this case, one can construct a quantum circuit with \(\mathcal O(\mathrm{polylog}\,N+l)\) gates and \(\mathcal O(\mathrm{polylog}\,N+l)\) ancilla qubits to implement this oracle \(O_D\).
\end{quote}
\noindent The rotation operators are simply controlled versions of single–qubit \(R_y(\theta )\) rotations, with the angle computed from the \(d'\) qubits in the control register. Using simple fixed-point arithmetic, such operators scale as \(\mathcal O (d')\).
\begin{table*}[bht]
    \centering
    \caption{\textbf{Quantum gate complexity and queries for 
    a single pipeline run.} Query complexity counts the number of calls to the primitive oracles \(O_L\) and \(\tilde{O}_X\).
    }
    \label{tab:qmmep_complexity}
    \resizebox{\columnwidth}{!}{
    \begin{tabular}{llccc}
        \toprule
        \textbf{Operator} & \textbf{Description} & \textbf{Gate Complexity} & \textbf{Qubit Count} & \textbf{Query Complexity} \\
        \midrule
        \(H\)            & elementary gate                  & \(1\) & \(1\) & -- \\
        \(O_L\)          & input model                      & \(\mathcal O(\text{polylog}\, N)\) & \(\mathcal O(\log N)\) & \(\mathcal O(1)\) \\
        \(\tilde O_X\)   & input model                      & \(\mathcal O(\text{polylog}\, N + d')\) & \(\mathcal O(\log N + d')\) & \(\mathcal O(1)\) \\
        \(\tilde O_K\)   & derived from \(O_L\)             & \(\mathcal O(\text{polylog}\, N + d')\) & \(\mathcal O(\log N + d')\) & \(\mathcal O(\log^2 s)\) \\
        \(F_X\)          & type of \(R_y\)                  & \(\mathcal O(d')\) & \(\mathcal O(d')\) & -- \\
        \(F_{K^{-1}}\)   & type of \(R_y\) (inverse)         & \(\mathcal O(d')\) & \(\mathcal O(d')\) & -- \\
        \(O_X\)          & \(1 \tilde O_X \cdot 5\, \mathrm{c}\text{-}F_X \cdot 1 \tilde O_X^T\) & \(\mathcal O(\text{polylog}\, N + d')\) & \(\mathcal O(\log N + d')\) & \(\mathcal O(1)\) \\
        \(O_{K^{-1}}\)   & \(1 \tilde O_K \cdot 1 F_{K^{-1}} \cdot 1 \tilde O_K^T\) & \(\mathcal O(\text{polylog}\, N + d')\) & \(\mathcal O(\log N + d')\) & \(\mathcal O(\log^2 s)\) \\
        \midrule
        \(\text{QMME}\)   & \text{embedding protocol} & \(\mathcal O(\text{polylog}\, N + d')\) & \(\mathcal O(\log N + d')\) & \(\mathcal O(\log^2 s)\) \\
        \text{cancer driver prediction pipeline} & \text{(\Cref{fig:qmme+swap-test})} & \(\mathcal O(\text{polylog}\, N +d')\) & \(\mathcal O(\log N + d')\) & \(\mathcal O(\log^2 s)\) \\
        \bottomrule
    \end{tabular}%
    }
\end{table*}

\noindent Recall that gene labels are predicted using a swap-test and two-qubit measurement statistics. In terms of gate count, the controlled-swap gate, that exchanges the training data and the test data, requires only \(\mathcal O(\log |V_l|)\) elementary gates to implement (Supplement~\cite{blank2020quantum}). The two-qubit measurements are repeated \(r\) times to estimate the expectation value \(\langle \sigma_z^{(a)} \sigma_z^{(l)}\rangle\), as defined in \Cref{eq:expectat-val-counts}. As such, the overall cost for this prediction scales linearly with \(r\), while the per-run complexities are those listed in~\Cref{tab:qmmep_complexity}. The statistical error of the estimate decreases with the number of repetitions \(r\) as \(\sqrt{r}\), following standard sampling statistics. As such, to get an estimate with error at most \(\epsilon\), the number of repetitions must scale as \(r=\mathcal O(1/\epsilon^2)\). 
\end{document}